\documentclass[journal]{IEEEtran}                                      
\usepackage{rotating}
\usepackage{amsmath}
\usepackage{graphicx,amssymb}
\usepackage{algorithm}
\usepackage{algorithmic}
\usepackage{amsthm,color}
\usepackage{diagbox}
\usepackage{multirow}
\usepackage{subcaption}
\usepackage{soul}

\makeatletter

\@addtoreset{algorithm}{section}
\makeatother

\newtheorem{theorem}{Theorem}[section]

\renewcommand{\theequation} {\thesection.\arabic{equation}}

\numberwithin{equation}{section}

\begin{document}
	\setstcolor{red}

\title{Distributed
algorithms to determine eigenvectors of matrices on spatially distributed networks}
\author{ Nazar Emirov,  Cheng Cheng, Qiyu Sun and Zhihua Qu 
\thanks{Nazar and Sun is with the  Department of Mathematics, University of Central Florida, Orlando, Florida 32816, USA;
Cheng is with the School of Mathematics, Sun Yat-Sen University,   Guangzhou, Guangdong, 510275, China;
 Qu is with the Department of Electrical and Computer Engineering, University of Central Florida, Orlando, Florida 32816, USA.
Emails: nazaremirov@knights.ucf.edu; chengch66@mail.sysu.edu.cn;   qiyu.sun@ucf.edu; Zhihua.Qu@ucf.edu.
This work is partially supported by
the  National Science Foundation (DMS-1816313).
}
}

\maketitle

\begin{abstract}
Eigenvectors of  matrices on a network have been used for understanding spectral clustering  and influence of a vertex.  
For matrices with small geodesic-width,
 we propose a distributed iterative algorithm in this letter to find eigenvectors associated with their given eigenvalues.
 We also consider the implementation of the proposed algorithm   at  the vertex/agent  level
in a spatially distributed network. 
\end{abstract}


\vskip-1.8mm

\section{Introduction}

Spatially distributed networks (SDNs) consist of  a large amount of agents, and each agent
 is equipped  with  subsystems for limited data processing and 
 direct communication link to its ``neighboring" agents
  within  communication  range.
SDNs  appear in  (wireless) sensor networks,
 smart grids, social network and many real world applications
\cite{Yick08}--\cite{Cheng17}.
In this letter, we describe the topological structure  of an SDN  by a  
finite graph
${\mathcal G}:=(V, E)$, 
and  its communication  range  $L$ by the maximal geodesic distance
such that   direct communication link between  agents $i,j\in V$ exists whenever
 $\rho(i,j)\le L$, where  the geodesic distance
 $\rho(i,j)$ is  the  number of edges in a shortest path connecting  $i, j\in V$.
   As SDNs do not have a central facility,
data processing on  SDNs
 should be designed at  the agent/vertex level with direct data exchanging between  neighboring vertices  in the communication range. 

 Matrices  on SDNs appear as  filters in graph signal processing,
  transition matrices in  Markov chains, state matrices in dynamic systems, 
  sensing matrices in sampling theory, and in many more applications  \cite{gleichsimareview2015, Cheng17}--\cite{cheng2020}.
   In the literature, their
   eigenspaces have been used to understand the
  communicability  between  vertices, spectral
clustering for the network and   influence of a vertex on the
network \cite{gleichsimareview2015,  langville2006}, \cite{Ma2010}--\cite{qu2017}.
 In this letter, we consider complex-valued matrices
 ${\bf A}=(A(i,j))_{i,j\in V}$ on the graph ${\mathcal G}=(V, E)$
 with limited geodesic-width $\omega({\bf A})$, which is the smallest nonnegative integer such that  $A(i, j)=0$ for all $i, j\in V$ satisfying $\rho(i, j)>\omega({\bf A})$.
 For a  matrix  
  with small geodesic-width, 
 we propose a distributed iterative algorithm
  to determine eigenvectors associated with its given eigenvalue, see Section \ref{sec:power_eigenvalues}. The proposed algorithm is based on the preconditioned gradient descent approach in \cite{cheng2020}
    for inverse filtering, and it can be implemented on  SDNs with communication range 
    larger than  geodesic-width of the matrix.
     Moreover, the  algorithm   
     has   its computational cost and communication expense
for subsystems equipped at every agent of the SDN being independent on the order of the graph ${\mathcal G}$.
In this letter, we also consider  finding principal eigenvectors associated with the minimal/maximal eigenvalue of a Hermitian matrix,
and  eigenvectors of a polynomial filter of graph shifts, see Sections \ref{sec:power_singularvalue} and  \ref{polynomialfilter.section}.

\vspace{-.06in}
\section{A distributed iterative  algorithm for  determining eigenvectors}\label{sec:power_eigenvalues}

\vspace{-.02in}

Let ${\mathcal G}=(V, E)$ be a connected, undirected and unweighted graph of order $N$.
Denote the set of all $s$-hop neighbors of a vertex $i\in V$  by $B(i,  s)=\{j\in V, \ \rho(j,i)\le s\}, s\ge 0$.
  For   a complex-valued matrix ${\bf A}=(A(i,j))_{i,j\in V}$  with small geodesic-width $ \omega({\bf A})$, we denote its Hermitian transpose by
  ${\bf A}^*$ and define
the diagonal preconditioning matrix ${\bf P}_{\bf A}$
 with  diagonal elements  
 \vspace{-.6em}\begin{eqnarray} \label{da.def}
&\hskip-0.08in  { P}_{\bf A}(i, i) := & \hskip-0.08in    \max_{k\in B(i, \omega({\bf A}))} \Big\{
 \max \Big( \sum_{j\in B(k, \omega({\bf A}))} |A(j, k)|, \nonumber\\
& \hskip.08in&\qquad
\sum_{j\in B(k, \omega({\bf A}))} |A(k, j)|\Big)\Big\},\ i\in V
 \vspace{-.6em}\end{eqnarray}
 \cite{cheng2020}.  In this section, we introduce a distributed iterative algorithm to find  eigenvectors of a  complex-valued matrix.

\begin{theorem}\label{maintheorem1.thm}
{\rm Let ${\bf A}$ be a complex-valued  matrix on the graph ${\mathcal G}$ of order $N$, 
${\bf P}_{\bf A}$ be the diagonal matrix  in  \eqref{da.def},
 and
${\bf Q}$  be a nonsingular diagonal matrix  such that
 \vspace{-.5em}\begin{equation}
 \label{maintheorem1.thm.eq1}
{\bf Q}- {\bf P}_{\bf A}\ {\rm is  \  positive\  semidefinite}.
 \vspace{-.5em}\end{equation}
Then for any initial 
${\bf x}_0\in {\mathbb C}^N$, 
the sequence ${\bf x}_n, n\ge 0$,  defined inductively by
 \vspace{-.6em}\begin{equation}\label{maintheorem1.thm.eq2}
{\bf x}_{n+1}= ({\bf I}- {\bf Q}^{-2} {\bf A}^* {\bf A}){\bf x}_{n},
 \vspace{-.5em} \end{equation}
 converges exponentially to
either the zero vector or an eigenvector associated with the zero eigenvalue of the matrix ${\bf A}$.
}\end{theorem}

The proof of Theorem \ref{maintheorem1.thm} will be given in Appendix \ref{proof.appendix}.

Take a  positive constant $c$ and define
a  diagonal matrix ${\bf Q}_c={\rm diag} (Q_c(i,i))_{i\in V}$ by
 \vspace{-.5em}
 \begin{equation}\label{distributedQii.def}
Q_c(i,i)=\max(P_{\bf A}(i, i), c), \ i\in V.
 \vspace{-.5em}\end{equation}
 Then ${\bf Q}_c$ is a nonsingular diagonal matrix satisfying
 \eqref{maintheorem1.thm.eq1}
 and it can be constructed at the vertex level, since
the preconditioning matrix ${\bf P}_{\bf A}$ can, see 
 \cite[Algorithm II.1]{cheng2020}.

Let   ${\bf H}=(H(i,j))_{i,j\in V}$  be a  matrix with small geodesic-width $\omega({\bf H})$ and $\lambda$ be its eigenvalue.
By selecting a random
 initial ${\bf x}_0$ with entries  i.i.d. 
 on $[0, 1]$, and applying the iterative algorithm
\eqref{maintheorem1.thm.eq2} to the matrix ${\bf A}={\bf H}-\lambda {\bf I}$ or $\lambda {\bf I}-{\bf H}$,
 we obtain from the proof of Theorem \ref{maintheorem1.thm} that the limit of the sequence ${\bf x}_n, n\ge 0$, is a nonzero vector (and hence
 an eigenvector  associated with the eigenvalue $\lambda$) almost surely.
Following the terminology in \cite{cheng2020}, we call the above algorithm to find eigenvectors
as a {\em preconditioned gradient descent algorithm}, PGDA for abbreviation.

The proposed PGDA is designed to implement distributedly  and synchronously at the vertex level, see Algorithm \ref{preconditioningmatrix.algorithm}.
For the implementation of  Algorithm \ref{preconditioningmatrix.algorithm}, 
 every vertex  $i\in V$ is required to have
the information of its $\omega({\bf H})$-hop neighbors,
 equipped direct communication link with its $\omega({\bf H})$-hop neighbors,
 and need  memory to store the eigenvalue $\lambda$, the iteration number $M$, the $i$-th diagonal entry of the  matrix ${\bf Q}$,
and entries $H(i,j)$ and  $H(j,i),j\in B(i,\omega({\bf H}))$
in the $i$-th row and column of the matrix ${\bf H}$.  Moreover, the computational cost and communication expense
for each vertex are independent on the order $N$ of the graph ${\mathcal G}$.
With the selection of the preconditioning matrix 
  as in \eqref{maintheorem1.thm.eq1}, we conclude that the proposed PGDA can be
 applied for an SDN with communication range $L$ to find eigenvectors associated with an {\bf  arbitrary} given eigenvalue for a matrix
 ${\bf H}$ with geodesic-width $\omega({\bf H})\le L$.

 \begin{algorithm}[t]
\caption{Realization of  the PGDA
at a vertex $i\in V$. }
\label{preconditioningmatrix.algorithm}
\begin{algorithmic}  

\STATE {\bf Inputs}: The total iteration number $M$,  the geodesic-width  $\omega({\bf H})$ of the matrix ${\bf H}=(H(i,j))_{i,j\in V}$,
 the set $B(i,\omega({\bf H}))$ of $\omega({\bf H})$-hop neighbors of the vertex $i$,
 the eigenvalue $\lambda$ of the matrix ${\bf H}$,
 entries $H(i,j)$ and $H(j, i), j \in B(i,\omega({\bf H}))$ in the $i$-th row and column
 of the matrix ${\bf H}$, and  the $i$-th diagonal entry $Q(i,i)$ of the matrix ${\bf Q}$.

\STATE{  
\bf Pre-iteration}: \ Compute $A(i, j)= H(i,j)-\lambda \delta(i,j)$
and $\tilde A(j,i)=(Q(i,i))^{-2} \big(\overline {H(j,i)}-\bar \lambda \delta(j,i)\big)$ for $j\in B(i, \omega({\bf H}))$, where $\delta$ is the Kronecker delta.
\STATE{  
\bf Initial}:
Select the $i$-th component $x_0(i)\in [0, 1]$ of the initial vector ${\bf x}_0$ randomly, and set  $n=0$.

\STATE{\bf Iteration}: \
  \begin{itemize}
    \item[{\bf  1.}] Send $x_n(i)$ to all neighbors $k\in B(i,\omega({\bf H}))\backslash \{i\}$ and receive $x_n(k)$ from neighbors $k\in B(i,\omega({\bf H}))\backslash \{i\}$.
\item [{\bf  2.}]  Evaluate $\tilde x_n(i)=\sum_{j\in B(i,\omega({\bf H}))} A(i,j) x_n(j)$.

\item [{\bf 3.}]  Send $\tilde x_n(i)$ to all neighbors $k\in B(i,\omega({\bf H}))\backslash \{i\}$ and receive $\tilde x_n(k)$ from neighbors $k\in B(i,\omega({\bf H}))\backslash \{i\}$.

    \item [{\bf 4.}] Evaluate $\widehat{ x}_n(i)=\sum_{j\in B(i,\omega({\bf H}))} \tilde A(j,i) \tilde x_n(j)$.

 \item[{\bf 5.}]  Set  $x_{n+1}(i)=x_n(i)-\widehat {x}_n(i)$ and $n=n+1$.

 \item [{\bf 6.}] return to step 1 if $n\le M$, go to Output otherwise.

 \end{itemize}
\STATE {\bf Output}: $ u(i)\approx x_{M}(i)$, where ${\bf u}=(u(i))_{i\in V}$ is the eigenvector.
\end{algorithmic} 
\end{algorithm}

For a left stochastic matrix on a network, principal eigenvectors  associated  with eigenvalue $1$
have positive entries by Perron-Frobenius theorem, and they
have been used
to determine the influence of a vertex, see \cite{gleichsimareview2015, langville2006} and references therein.
Let ${\bf W}=(W(i,j))_{i,j\in V}$ be the hyperlink matrix on a network described by a graph ${\mathcal G}=(V, E)$, where weights
$W(i,j)=0$ for $(i,j)\not\in E$  and
$W(i,j)=1/d_j$ for $(i,j)\in E$, the reciprocal of the degree $d_j$  of a vertex $j$. The matrix ${\bf W}$ is a left stochastic matrix with
$1$ as the leading eigenvalue.
Applying the proposed  PGDA to the hyperlink matrix ${\bf W}$, we can locally evaluate  principal eigenvectors of the hyperlink matrix and hence identify the local influence of a vertex on its neighborhood.


\section{Principal eigenvectors of  Hermitian matrices}\label{sec:power_singularvalue}

In this section, we consider finding eigenvectors
 associated with the minimal/maximal eigenvalue
of a Hermitian matrix
on a  graph  ${\mathcal G}=(V, E)$ of order $N$ 
in a distributed manner.

\begin{theorem}\label{symmetricmain.thm}
{\rm Let ${\bf A}=(A(i,j))_{i,j\in V}$ be a positive semidefinite matrix on the graph ${\mathcal G}$ with its geodesic-width 
$\omega({\bf A})$,
and  ${\bf Q}^{\rm sym}={\rm diag}(Q^{\rm sym}(i,i))_{ i\in V}$ be a nonsingular diagonal matrix  
 satisfying
  \vspace{-.5em}
\begin{equation} \label{symmetricmain.thm.eq1}
Q^{\rm sym}(i,i)\ge \sum_{j\in B(i, \omega({\bf A}) )}|A(i,j)|,\  i\in V.
 \vspace{-.4em} \end{equation}
Then for any ${\bf x}_0\in \mathbb C^N$,   the sequence ${\bf x}_n, n\ge 1$, defined by
 \vspace{-.4em} \begin{equation}\label{symmetricmain.thm.eq2}
{\bf x}_{n+1}= ({\bf I}- ({\bf Q}^{\rm sym})^{-1} {\bf A}) {\bf x}_n, 
 \vspace{-.4em} \end{equation}
 converges exponentially to
either the zero vector or an eigenvector associated with the zero eigenvalue of the matrix ${\bf A}$.
}
\end{theorem}
 \vspace{-.3em}

 \begin{IEEEproof}
 Following the argument in \cite[Theorem III.1]{cheng2020} and applying \eqref{symmetricmain.thm.eq1},
we  obtain that
${\bf Q}^{\rm sym}-{\bf A}$  
is positive semidefinite.
This together with the positive semidefiniteness of the matrix ${\bf A}$ implies that
all eigenvalues of the Hermitian matrix ${\bf B}^{\rm sym}:={\bf I}-({\bf Q}^{\rm sym})^{-1/2}{\bf A}({\bf Q}^{\rm sym})^{-1/2}$
are in the unit interval $[0, 1]$, cf. \eqref{maintheorem1.thm.pfeq3} in Appendix \ref{proof.appendix}.
Applying similar argument used in the proof of Theorem \ref{maintheorem1.thm} with ${\bf Q}$ and ${\bf A}^*{\bf A}$ replaced by
$({\bf Q}^{\rm sym})^{1/2}$ and ${\bf A}$ respectively, we obtain
 \vspace{-.3em}
\begin{equation}\label{symmetricmain.thm.pfeq1}
 \|({\bf Q}^{\rm sym})^{1/2}({\bf x}_n-{\bf u})\|_2\le \|({\bf Q}^{\rm sym})^{1/2}{\bf x}_0\|_2  r^n,\ \ n\ge 0
  \vspace{-.3em}\end{equation}
for  some vector ${\bf u}\in {\mathbb C}^N$, where $r$ is  the largest eigenvalue of ${\bf B}^{\rm sym}$ in $[0, 1)$.
This together with the nonsingularity of the  matrix ${\bf Q}^{\rm sym}$ proves the exponential convergence of
${\bf x}_n, n\ge 0$.

Taking limit in \eqref{symmetricmain.thm.eq2} proves
${\bf A}{\bf u}={\bf 0}$, and hence completes the proof.
  \end{IEEEproof}

Let ${\bf H}$ be a Hermitian matrix  with minimal eigenvalue $\lambda_{\min}$ and maximal eigenvalue $\lambda_{\max}$.
 Then ${\bf A}_1={\bf H}-\lambda_{\min}{\bf I}$ and ${\bf A}_2=\lambda_{\max}{\bf I} - {\bf H}$ have eigenvalue zero and they are positive semidefinite. Then applying the iterative algorithm \eqref{symmetricmain.thm.eq2} to ${\bf A}_1$ (resp. ${\bf A}_2$)
 with a random initial ${\bf x}_0$  having entries  i.i.d on $[0, 1]$,
  we obtain the principal eigenvectors associated with minimal (resp. maximal) eigenvalues of the Hermitian matrix ${\bf H}$ by  Theorem \ref{symmetricmain.thm}.

For a positive semidefinite matrix ${\bf A}=(A(i,j))_{i,j\in V}$  with  geodesic-width $\omega({\bf A})$,
a nonsingular diagonal matrix ${\bf Q}_c^{\rm sym}={\rm diag}(Q_c^{\rm sym}(i,i))_{ i\in V}$ satisfying
 \eqref{symmetricmain.thm.eq1} can be constructed  at the vertex level by setting
 \vspace{-.4em}
\begin{equation}\label{distributedQsymmetricii.def}
Q_c^{\rm sym}(i,i)=\max\Big(\sum_{j\in B(i, \omega({\bf A}) )}|A(i,j)|, c\Big), \ i\in V,
\vspace{-.4em} \end{equation}
where $c$ is a positive constant, cf. \eqref{distributedQii.def}. With the above selection of the preconditioning matrix in \eqref{symmetricmain.thm.eq2}, we can find  eigenvectors associated with minimal/maximal eigenvalues of a
Hermitian matrix  by the distributed iterative algorithm \eqref{symmetricmain.thm.eq2} implementable at the vertex level,
 see Algorithm
\ref{Qsym.algorithm}.
  Following the terminology in \cite{cheng2020}, we call the  algorithm  \eqref{symmetricmain.thm.eq2} with a random initial
  having entries i.i.d on $[0, 1]$
as a {\em symmetric preconditioned gradient descent algorithm}, SPGDA for abbreviation.
Comparing with  Algorithm \ref{preconditioningmatrix.algorithm} to find eigenvectors of an  arbitrary matrix,
the  Algorithm \ref{Qsym.algorithm} to find principal eigenvectors of a Hermitian matrix
has less computational cost and
communication expense
 in each iteration. Our numerical simulations
in  Section \ref{sec:power_polynomial} also indicate that it may have faster convergence.

 \begin{algorithm}[t]
\caption{Realization of the  SPGDA at a vertex $i\in V$. }
\label{Qsym.algorithm}
\begin{algorithmic}  

\STATE {\bf Inputs}: The total iteration number $M$, 
  the geodesic-width $\omega({\bf A})$ of the positive semidefinite matrix ${\bf A}$,
 the set $B(i,\omega({\bf A}))$ of $\omega({\bf A})$-hop neighbors of the vertex $i$,
 entries $A(i,j), j \in B(i,\omega({\bf A}))$ in the $i$-th row
 of the matrix ${\bf A}$ and the $i$-th entry $Q^{\rm sym}(i,i)$ of the diagonal matrix ${\bf Q}^{\rm sym}$.

\STATE{  
\bf Pre-iteration}: \ Evaluate $\tilde A(i, j)=(Q^{\rm sym}(i,i))^{-1} A(i,j)$, $j\in B(i, \omega({\bf A}))$.
\STATE{  
\bf Initial}:
Select 
 $x_0(i)$ randomly in $[0, 1]$, 
and set  $n=0$.

\STATE{\bf Iteration}: \
  \begin{itemize}
    \item[{\bf  1.}] Send $x_n(i)$ to all neighbors $k\in B(i,\omega({\bf A}))\backslash \{i\}$ and receive $x_n(k)$ from neighbors $k\in B(i,\omega({\bf A}))\backslash \{i\}$.
\item [{\bf  2.}]  Evaluate $ x_{n+1}(i)=x_n(i)-\sum_{j\in B(i,\omega({\bf A}))} \tilde A(i,j) x_n(j)$ and set $n=n+1$.

 \item [{\bf 3.}] return to step 1 if $n\le M$, go to Output otherwise.

 \end{itemize}
\STATE {\bf Output}: $ u(i)\approx y_{M}(i)$, where ${\bf u}=(u(i))_{i\in V}$. 
\end{algorithmic}
\vspace{-.03in}
\end{algorithm}

\vspace{-.08in}
\section{Eigenvectors of polynomial filters}
\label{polynomialfilter.section}

  Graph filter is a  fundamental concept in graph signal
processing  and it has been
used in many applications such as  denoising and consensus of multi-agent systems
\cite{shuman13,   Ortega18, Waheed18, shuman18, Emirov20, cheng2020,  hammod11}--\cite{jiang19}.
An	elementary graph filter  is a {\em graph shift}, which has  $1$  as its geodesic-width. Graph filters
 in most of literature  are designed to be polynomials
   \vspace{-0.6em}\begin{equation}\label{MultiShiftPolynomial}
	{\bf A}=h({\bf S}_1, \ldots, {\bf S}_d)=\sum_{ l_1=0}^{L_1} \cdots \sum_{ l_d=0}^{L_d}  h_{l_1,\dots,l_d}{\bf S}_1^{l_1}\cdots {\bf S}_d^{l_d}
	\vspace{-0.6em}\end{equation}
 of commutative graph
 shifts ${\bf S}_1,...,{\bf S}_d$, i.e., ${\bf S}_k{\bf S}_{k'}={\bf S}_{k'}{\bf S}_k$ for all $1\le k,k'\le d$,
  where the multivariate polynomial
$h(t_1, \ldots, t_d)=\sum_{ l_1=0}^{L_1} \cdots \sum_{ l_d=0}^{L_d}  h_{l_1,\dots,l_d} t_1^{l_1} \ldots t_d^{l_d}$
 has polynomial coefficients $h_{l_1,\dots,l_d}$, $0\le l_k\le L_k, 1\le k\le d$
 \cite{Leus17}--\cite{mario19}, \cite{jiang19}--\cite{ Emirov19}.
 On the graph ${\mathcal G}=(V, E)$, a polynomial filter ${\bf A}$ in \eqref{MultiShiftPolynomial}
 can be represented by a matrix ${\bf A} = (A(i,j))_{i,j\in V}$, which
has geodesic-width no more than the degree of the polynomial $h$, i.e., $\omega({\bf A})\le \sum_{k=1}^d L_k$.
Then we can apply the PGDA (resp. the SPGDA if ${\bf A}$ is Hermitian) to find eigenvectors associated with any given eigenvalue (resp.
the minimal/maximal eigenvalues) on SDNs with communication range $L\ge \sum_{k=1}^d L_k$.
In this section,  we propose iterative  algorithms to determine eigenvectors associated with a polynomial graph filter, which can be implemented on an SDN with {\bf  $1$} as its communication range, i.e., direct communication exists between all adjacent vertices.  

Observe that
 \vspace{-0.4em}
\begin{equation} \label{MultiShiftPolynomial.polynomial2}
{\bf A}^*=\sum_{ l_1=0}^{L_1} \cdots \sum_{ l_d=0}^{L_d} \overline{ h_{l_1,\dots,l_d}} ({\bf S}_d^*)^{l_d}\cdots ({\bf S}_1^*)^{l_1}
 \vspace{-0.4em}\end{equation}
is a polynomial graph filter of commutative shifts ${\bf S}_1^*,...,{\bf S}_d^*$.
Then applying Algorithm II.2  in  \cite{Emirov20} to implement the filtering procedure associated with polynomial graph filters ${\bf A}$ and ${\bf A}^*$, we can
implement  each iteration in the PGDA \eqref{maintheorem1.thm.eq2} and the SPGDA \eqref{symmetricmain.thm.eq2} in finite steps with each step including  data exchanging between  adjacent vertices only, see Algorithm \ref{onehopeigenvalue.algorithm} to determine eigenvectors associated with eigenvalue zero. This concludes that
eigenvectors for a polynomial graph filter 
 on SDNs with communication range  $1$  can be obtained by applying Algorithm \ref{onehopeigenvalue.algorithm} in each iteration.

\begin{algorithm}[t]
\caption{Realization of  each iteration in the iterative algorithms  \eqref{maintheorem1.thm.eq2} and  \eqref{symmetricmain.thm.eq2} at a vertex $i\in V$ for a polynomial filter ${\bf A}$. }
\label{onehopeigenvalue.algorithm}
\begin{algorithmic}  

\STATE {\bf Inputs}: Polynomial coefficients $h_{l_1,\ldots, l_d}, 0\le l_1\le L_1, \ldots, 0\le l_d\le L_d$ of the polynomial filter ${\bf A}$ in \eqref{MultiShiftPolynomial}, the set ${\mathcal N}_i$  
of all adjacent vertices $j$ of the vertex $i$,
entries   $S_k(i,j)$ and $S_k(j,i), j\in {\mathcal N}_i$ 
of
			graph shifts ${\bf S}_k, 1\le k\le d$, the $i$-th diagonal entry $Q(i,i)$ of the matrix ${\bf Q}$,  
and
  the $i$-th entry $x_{n}(i)$ of the input vector ${\bf x}_{n}=(x_{n}(k))_{k\in V}$ at $n$-th iteration,

\STATE{\bf 1}: \  Apply Algorithm II.2  in  \cite{Emirov20} to implement the polynomial filter procedure ${\bf x}\longmapsto {\bf A}{\bf x}$
at the vertex $i$.
The input is  the $i$-th entry $x_n(i)$ of ${\bf x}_n$   and
the output is  the $i$-th entry $\hat x_n(i)$
 of  $\hat {\bf x}_n={\bf A}{\bf x}_n 
 =:(\hat x_n(k))_{k\in V}$.

\STATE{\bf 2}: \  Apply Step 1 with the matrix ${\bf A}$ replaced by its complex conjugate ${\bf A}^*$ and
the input $x_n(i)$ by  $\hat x_n(i)$. 
 The output is   the $i$-th entry $\check{x}_n(i)$ of the vector $\check{\bf x}_n= {\bf A}^*\hat {\bf x}_n 
 =:(\check{x}_n(k))_{k\in V}$.

\STATE{\bf 3}: \ Evaluate $x_{n+1}(i)=x_{n}(i)- (Q(i,i))^{-2} \check{x}_n(i) $
and $\tilde x_{n+1}(i)=x_{n}(i)- (Q(i,i))^{-1} \hat{x}_n(i)$.

\STATE {\bf Outputs}: The outputs  $x_{n+1}(i)$ and $\tilde x_{n+1}(i)$ are  the $i$-th entry of  ${\bf x}_{n+1}$
 at $n$-th iteration in \eqref{maintheorem1.thm.eq2} and  \eqref{symmetricmain.thm.eq2} respectively.
\end{algorithmic}
\vspace{-.03in}
\end{algorithm}

Now it remains to construct diagonal
matrices satisfying \eqref{maintheorem1.thm.eq1} and \eqref{symmetricmain.thm.eq1} on SDNs with communication range $1$.
For the polynomial graph filter ${\bf A}$ in \eqref{MultiShiftPolynomial}, define
diagonal matrices
$\widehat {\bf Q}_c={\rm diag} (\widehat {Q}_c(i,i))_{i\in V}$ and ${\widehat {\bf Q}}_c^{\rm sym}={\rm diag} ({\widehat {Q}}^{\rm sym}_c(i,i))_{i\in V} $ by
\begin{equation}\label{Rprecondition.def}
\vspace{-0.6em}
\hskip-.15in\widehat Q_c(i,i)= \max_{\rho(j,i)\le L} \max\Big\{\sum\limits_{k\in  V}\widehat A(j,k),\sum\limits_{k\in V}\widehat A(k,j), c\Big\}\hskip-.05in
\vspace{-0.4em}\end{equation}
 and
\begin{equation}\label{Rpreconditionsym.def}
\vspace{-0.6em} {\widehat Q}_c^{\rm sym}(i,i)= \max\Big\{\sum\limits_{k\in  V}\widehat A(j,k), c \Big\}, \ i\in V,
\vspace{-0.2em} \end{equation}
 where $c$ is a positive number, $|{\bf S}_k|=(|S_k(i,j)|)_{i,j\in V}, 1\le k\le d$, and
 \vspace{-0.8em}
 \begin{equation*}\label{MultiShiftPolynomial2}
	( \widehat A(i,j))_{i,j\in V}=:\widehat {\bf A}:=\sum_{ l_1=0}^{L_1} \cdots \sum_{ l_d=0}^{L_d}  |h_{l_1,\dots,l_d}| |{\bf S}_1|^{l_1}\cdots |{\bf S}_d|^{l_d}.
 \vspace{-0.6em}
\end{equation*}
 One may verify that
$|A(i,j)|\le \widehat A(i,j)$ for all $i,j\in V$.
%
Therefore the  matrices  $\widehat {\bf Q}_c$   in
\eqref{Rprecondition.def} and ${\widehat {\bf Q}}_c^{\rm sym}$
in \eqref{Rpreconditionsym.def}
satisfy  \eqref{maintheorem1.thm.eq1}
and
\eqref{symmetricmain.thm.eq1} respectively. Moreover,
as shown in Algorithm \ref{onehopconstruction.algorithm}, they can be
constructed  at the vertex level  in finite steps  such that in  each step,  each vertex needs  to exchange data with adjacent vertices only.

\begin{algorithm}[t]
\caption{Construction of  diagonal entries $\widehat Q_c(i, i)$ and
${\widehat Q}_c^{\rm sym}(i, i)$
at a vertex  $i\in V$ for a polynomial filter ${\bf A}$. }
\label{onehopconstruction.algorithm}
\begin{algorithmic}  

\STATE {\bf Inputs}: The positive constant $c$,
polynomial coefficients $h_{l_1,\ldots, l_d}, 0\le l_1\le L_1, \ldots, 0\le l_d\le L_d$, of the polynomial filter ${\bf A}$, entries   $S_k(i,j)$ and $S_k(j,i)$ for all $1\le k\le d$ and $j\in {\mathcal N}_i$, the set of all adjacent vertices  of the vertex $i$.

\STATE{\bf 1}: \  Apply Algorithm II.2  in  \cite{Emirov20} to implement the polynomial filter procedure ${\bf 1}\longmapsto \widehat {\bf A}{\bf 1}$
at the vertex $i$.
The input is  the $i$-th entry $1$ of the all-one vector ${\bf 1}$  and
the output is  the $i$-th entry  $a_1(i)$
 of the vector $\widehat {\bf A}{\bf 1}=:(a_1(k))_{k\in V}$.

\STATE{\bf 2}: \
Apply Step 1 with the same input but the  filter $\widehat {\bf A}$ replaced by $\widehat {{\bf A}^*}$.
The output is  the $i$-th entry
 $a_2(i)$ of the vector $\widehat {{\bf A}^*}{\bf 1}=:(a_2(k))_{k\in V}$.

\STATE{\bf 3}: \ Evaluate $q_0(i)=\max (a_1(i), a_2(i), c) $ and set $l=0$.

\STATE {{\bf 4}: {\bf Finite-step iteration}}: \
  \begin{itemize}
    \item[{\bf  4a)}] Send $q_l(i)$ to all adjacent vertices  $k\in {\mathcal N}_i$ and receive $q_l(k)$ from all adjacent vertices $k\in {\mathcal N}_i$. 
\item [{\bf  4b)}]  Compare $q_l(i)$ with  $q_l(k), k\in {\mathcal N}_i$ and define $q_{l+1}(i)=\max(q_l(i), \max_{k\in {\mathcal N}_i} q_l(k))$ and set $l:=l+1$.

 \item [{\bf 4c)}] Return to step 1 if $l\le L_1+\ldots+L_d$, go to Outputs otherwise.

 \end{itemize}

\STATE {\bf Outputs}: $\widehat Q_c(i, i)=q_L(i)$ and
${\widehat Q}_c^{\rm sym}(i, i)=\max(a_1(i), c)$.
\end{algorithmic}\vspace{-.03in}
\end{algorithm}

\vspace{-.1in}
\section{Numerical Simulations}\label{sec:power_polynomial}

\begin{figure}[t]
\begin{center}

\includegraphics[width=28mm, height=20mm]{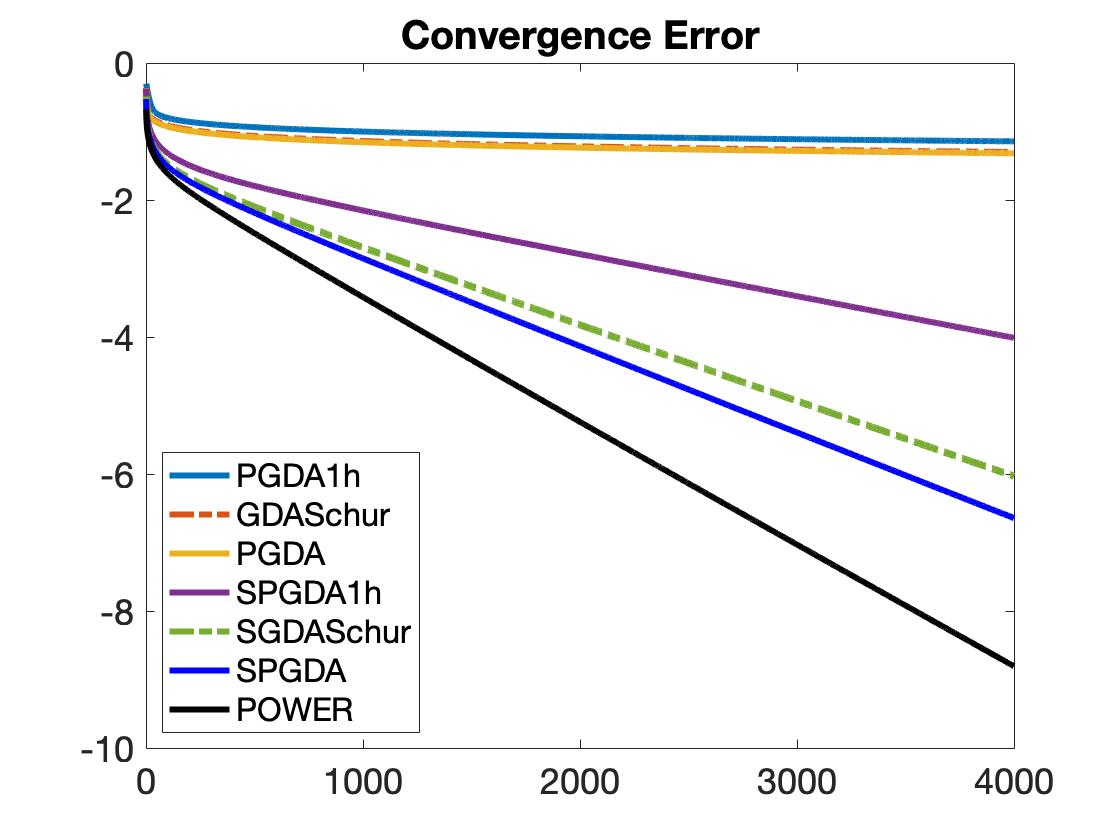}
\includegraphics[width=28mm, height=20mm]{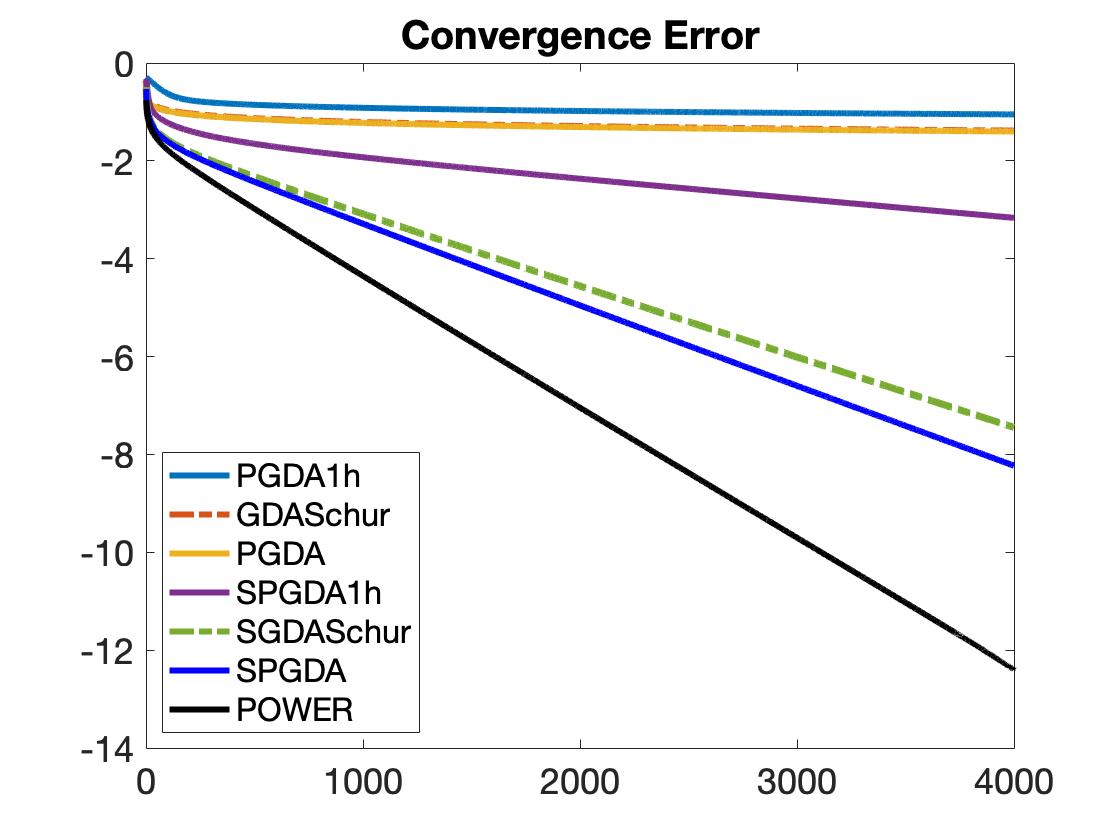}
\includegraphics[width=28mm, height=20mm]{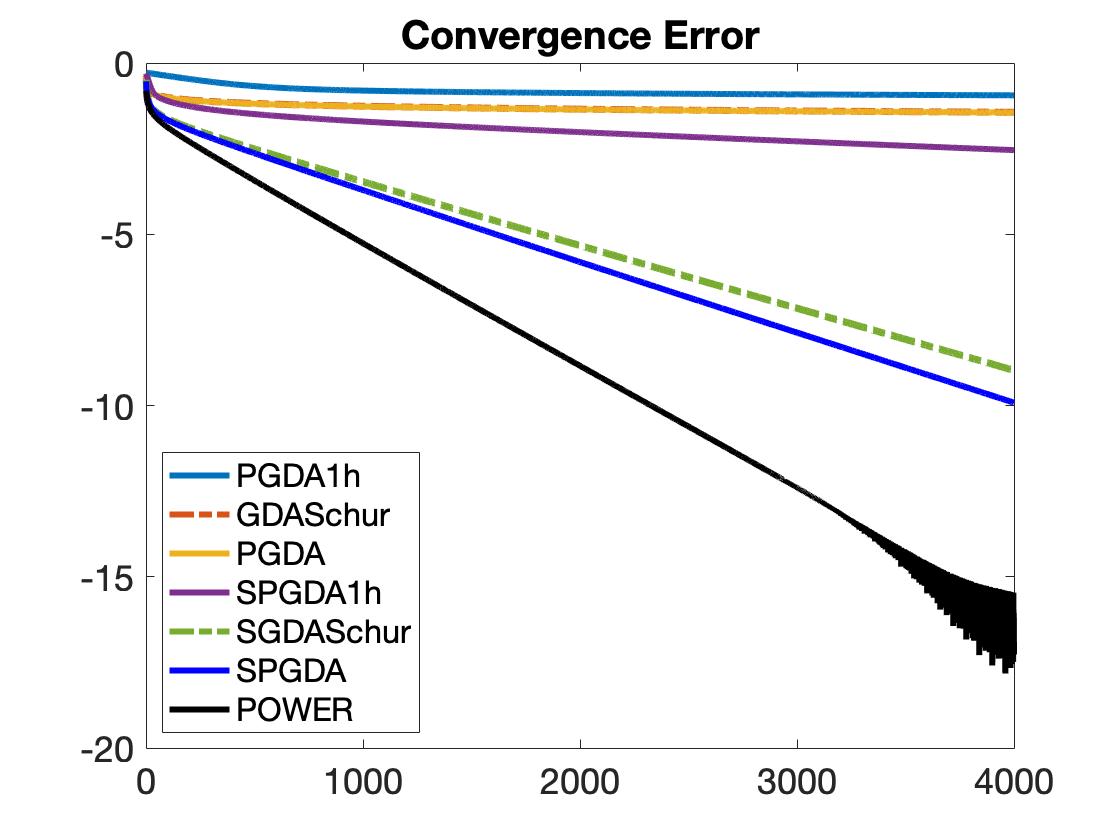}
\\
\includegraphics[width=28mm, height=20mm]{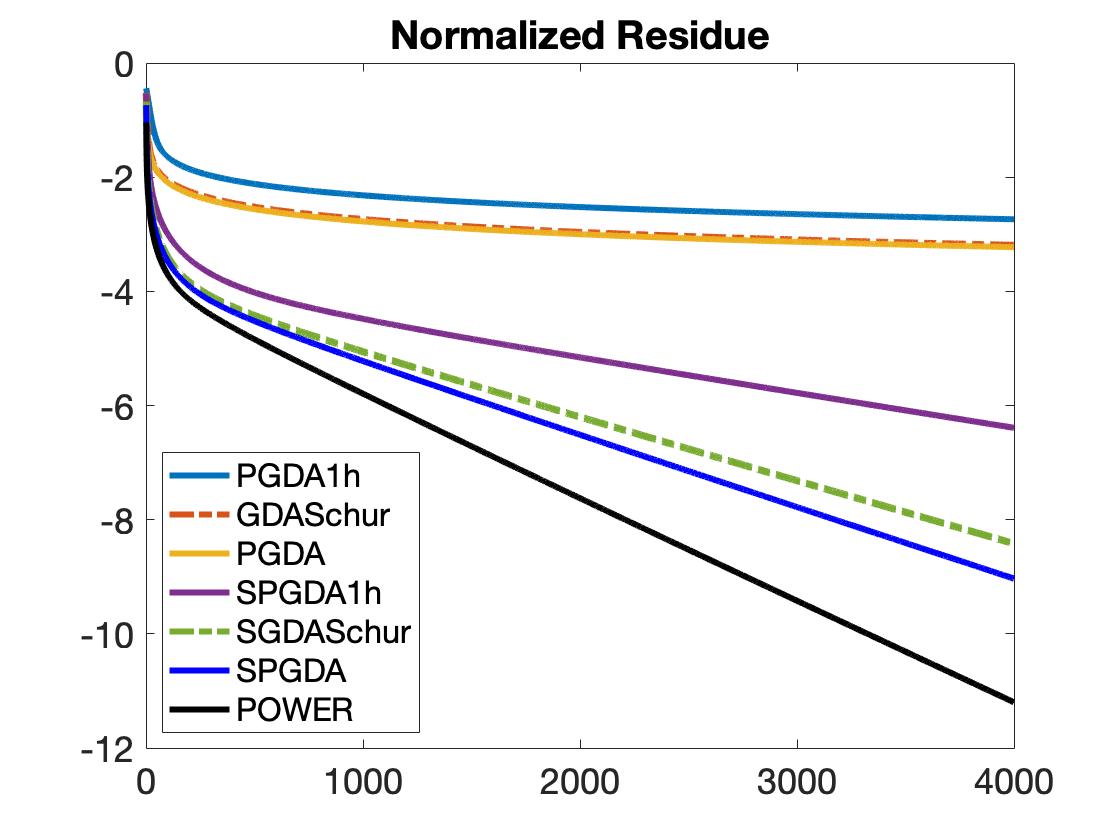}
\includegraphics[width=28mm, height=20mm]{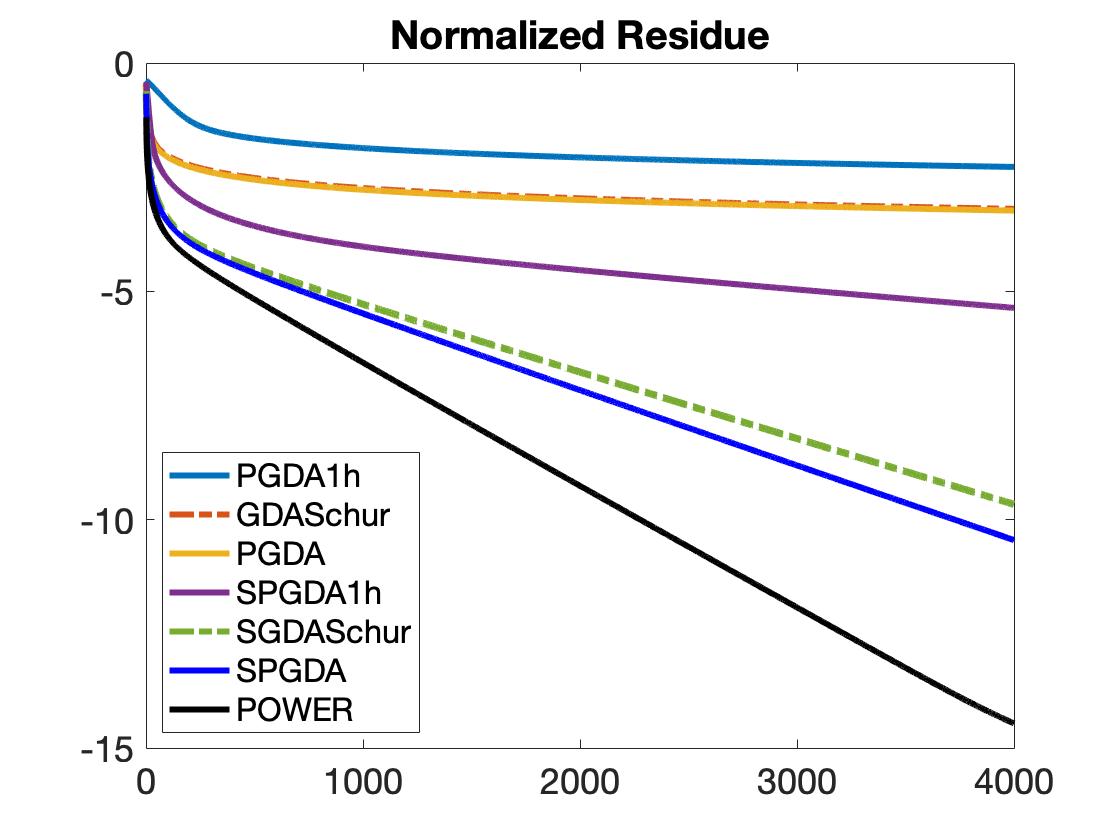}
\includegraphics[width=28mm, height=20mm]{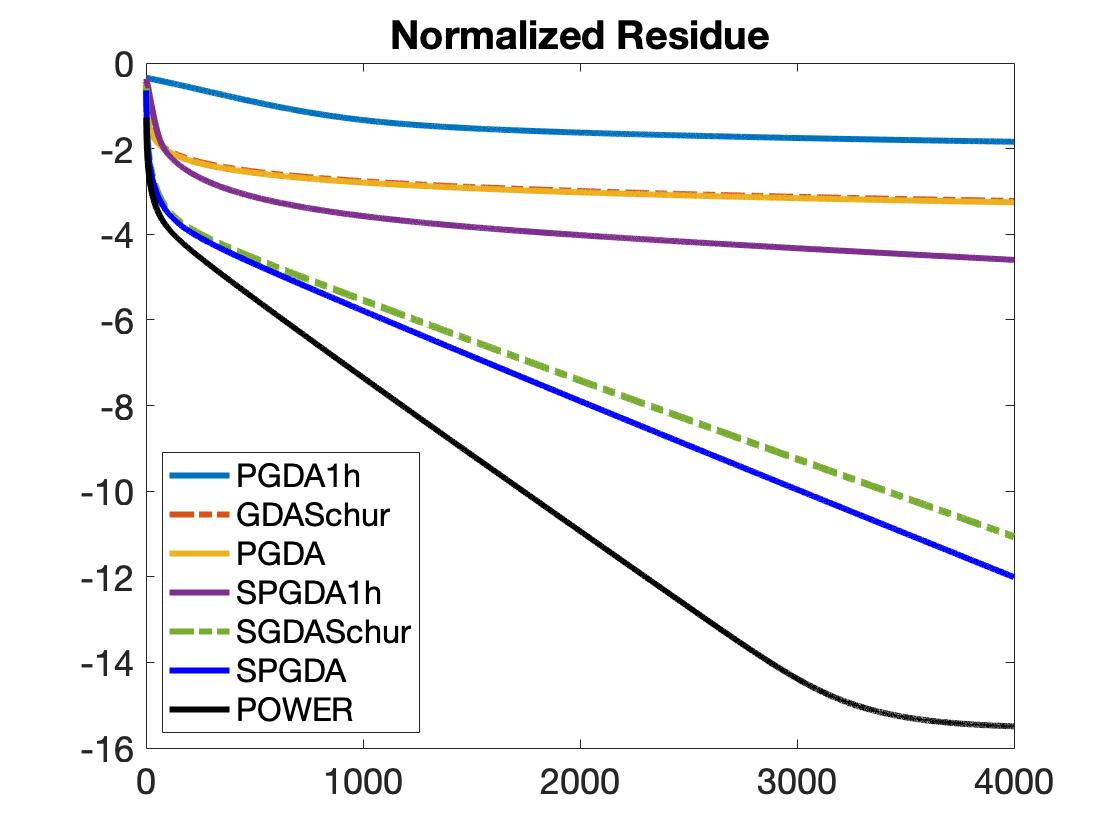}
\caption{ 
Plotted on the first and second rows are
average over 500 trials of  the convergence  errors ${\rm CE}(n)$ and  normalized residues ${\rm RE}(n), 1\le n\le 4000$,
in the
logarithmic scale, while from left to right
are lowpass spline filters ${\bf H}_{0, m}^{\rm spln}$ of orders $m=2, 3, 4$ on the random geometric graph ${\mathcal G}_{512}$. 
  }
\label{fig2}
\end{center}
\vspace{-.5em}
\end{figure}

Let ${\mathcal G}_{N}=(V_{N}, E_{N}), N\ge 2$, be random geometric graphs
  with $N$ vertices 
deployed on  
 $[0, 1]^2$ and  an undirected edge between two vertices in $V_N$ existing if their physical
distance is not larger than
$\sqrt{2/N}$ 
\cite{jiang19, Nathanael2014}.
In this section, we consider finding eigenvectors associated with maximal eigenvalue $1$ of
lowpass spline filters ${\bf H}_{0, m}^{\rm spln}=({\bf I}-{\bf L}^{\rm sym}/2)^m, m\ge 1$,
where ${\bf L}^{\rm sym}$ is the symmetric normalized Laplacian matrix on
the graph ${\mathcal G}_N$
  \cite{jiang19,dragotti2017}. 
In the simulations, we take $c=0.01$ and  use PGDA and PGDA1h to denote the
PGDA with ${\bf A}$ replaced by ${\bf I}-{\bf H}_{0, m}^{\rm spln}$ and  ${\bf Q}$  by   ${\bf Q}_c$ in \eqref{distributedQii.def} and $\widehat {\bf Q}_c$
in \eqref{Rprecondition.def}
 respectively, and similarly we use SPGDA and SPGDA1h to denote the SPGDA with ${\bf A}$ replaced by ${\bf I}-{\bf H}_{0, m}^{\rm spln}$ and  ${\bf Q}$ by
 ${\bf Q}_c^{\rm sym}$ in  \eqref{distributedQsymmetricii.def} and  $\widehat{\bf Q}_c^{\rm sym}$  in \eqref{Rpreconditionsym.def}
 respectively. 
For the sequences
  ${\bf x}_n, n\ge 0$, in the PGDA, SPGDA, PGDA1h and SPGDA1h and their limits ${\bf u}$, define
convergence errors
${\rm CE}(n)={\rm log}_{10}\|\tilde {\bf x}_{n}-\tilde {\bf u} \|_2$
 and normalized residues
${\rm NR}(n)={\rm log}_{10}\|({\bf I}-{\bf H}_{0, m}^{\rm spln}) \tilde {\bf x}_{n}\|_2, n\ge 0,$
in the logarithmic scale, where  $\tilde {\bf x}_{n}=  {\bf x}_n /\|{\bf x}_n\|_2$, $\tilde {\bf u}={\bf u}/\|{\bf u}\|_2$, and
 $\|{\bf x}\|_2=\big(\sum_{j\in V} |x(j)|^2\big)^{1/2}$ for  ${\bf x}=(x_j)_{j\in V}$.
Shown in
Figure \ref{fig2} are  the average of convergence errors
${\rm CE}(n)$
and normalized residues
${\rm RE}(n), n\ge 0,$
 over 500 trials. This demonstrates the exponential convergence
  of the sequence ${\bf x}_n, n\ge 0$,  in the proposed distributed iterative algorithms 
 to eigenvectors
 associated with eigenvalue $1$ of lowpass spline filters, 
 which is proved in
   Theorems \ref{maintheorem1.thm} and \ref{symmetricmain.thm}.

 For a matrix ${\bf A}$ on a graph ${\mathcal G}=(V, E)$, define its Schur norm  by
  $\|{\bf A}\|_{\mathcal S}=\max_{i\in V} P_{\bf A}(i,i)$, where $P_{\bf A}(i,i), i\in V$, are given by \eqref{da.def}
  \cite{Cheng17, cheng2020}.
 For the case that the constant $c$
  in
  \eqref{distributedQii.def} and \eqref{distributedQsymmetricii.def}  is so chosen that
$c\ge \|{\bf A}\|_{\mathcal S}$,
 the  preconditioning matrices ${\bf Q}_c$  and
 ${\bf Q}_c^{\rm sym}$ become a multiple of the identity matrix ${\bf I}$ and
 the corresponding PGDA and SPGDA are
the conventional gradient descent algorithm  
and its symmetric version
respectively \cite{Leus17,  shuman18, Emirov20, cheng2020, sihengTV15}.
We denote the above
 algorithms with $c=\|{\bf A}\|_{\mathcal S}$
by GDASchur and SGDASchur respectively, see  Figure \ref{fig2} for their performance.
  Since  $1$  is the maximal eigenvalue of matrices ${\bf H}_{0, m}^{\rm spln}, m\ge 1$,
we can use
 the conventional power iteration method  with entries of the initial ${\bf x}_0$ randomly selected in $[0, 1]$,  POWER for abbreviation,
  to find principal eigenvectors   
   \cite{golubbook2013}.
Presented in
Figure \ref{fig2} is its performance.
From Figure \ref{fig2}, we observe that
the centralized algorithm POWER  has fastest convergence 
 to find eigenvectors 
  of  matrices ${\bf H}_{0, m}^{\rm spln}, 2\le m\le  4$,
as followed are   the distributed  algorithm SPGDA, the centralized algorithm SPGDASchur and the distributed  algorithm SPGDA1h,
the next are the distributed algorithm PGDA  and the centralized algorithm  GDASchur, and the distributed algorithm PGDA1h has slowest convergence.

\begin{appendices}
\renewcommand{\thesectiondis}[2]{\Alph{section}:}
\section{Proof of Theorem \ref{maintheorem1.thm}}
\label{proof.appendix}

By nonsingularity of the  matrix ${\bf Q}$,  it suffices 
to prove
\begin{equation}\label{maintheorem1.thm.pfeq0}
 \|{\bf Q}({\bf x}_n-{\bf u})\|_2\le \|{\bf Q}{\bf x}_0\|_2  r^n,\ \ n\ge 0\end{equation}
for some ${\bf u}$ satisfying ${\bf A}{\bf u}={\bf 0}$, where $r\in (0, 1)$.

 Set ${\bf B}={\bf I}- {\bf Q}^{-1}{\bf A}^* {\bf A} {\bf Q}^{-1}$ and
let  ${\bf u}_i$ be orthonormal eigenvectors associated with eigenvalues $\gamma_i$ of the Hermitian matrix ${\bf B}$ that
satisfy 
\begin{equation} \label{maintheorem1.thm.pfeq2}
{\bf B} {\bf u}_i=\gamma_i {\bf u}_i,\ \  1\le i\le N.
\end{equation}
Following the argument in \cite[Theorem II.1]{cheng2020} and applying \eqref{maintheorem1.thm.eq1},
we obtain that
${\bf Q}^2-{\bf A}^*{\bf A}$ is positive semidefinite.
This together with nonsingularity of the matrix ${\bf Q}$ implies that
\begin{equation} \label{maintheorem1.thm.pfeq3}
0\le \gamma_i\le 1, \ 1\le i\le N.
\end{equation}

Write
$ {\bf Q} {\bf x}_0=\sum_{i=1}^N \langle {\bf Q} {\bf x}_0, {\bf u}_i\rangle  {\bf u}_i$,
where  $\langle \cdot, \cdot\rangle$ is the standard inner product on ${\mathbb C}^N$. 
By \eqref{maintheorem1.thm.eq2}, we have that
${\bf Q} {\bf x}_n={\bf B} {\bf Q} {\bf x}_{n-1}, \  n\ge 1$.
Therefore 
 \vspace{-.5em}
\begin{equation} \label{maintheorem1.thm.pfeq5}
{\bf Q}{\bf x}_n= {\bf B}^n {\bf Q}{\bf x}_0=\sum_{i=1}^N \gamma_i^n \langle {\bf Q} {\bf x}_0, {\bf u}_i\rangle  {\bf u}_i, \   n\ge 0.
 \vspace{-.5em} \end{equation}
Define ${\bf u}=\sum_{\gamma_i=1} \langle {\bf Q} {\bf x}_0, {\bf u}_i\rangle  {\bf Q}^{-1}{\bf u}_i$.
Then  by  \eqref{maintheorem1.thm.pfeq3}, \eqref{maintheorem1.thm.pfeq5}
and the orthonormality of ${\bf u}_i, 1\le i\le N$, we obtain
 \vspace{-.5em}
 \begin{eqnarray}
 \hskip-0.08in \|{\bf Q}({\bf x}_n-{\bf u})\|_2 & \hskip-0.08in = & \hskip-0.08in
\Big(\sum_{0\le \gamma_i<1} |\langle {\bf Q} {\bf x}_0, {\bf u}_i\rangle |^2 \gamma_i^{2n}\Big)^{1/2}\nonumber\\
 \hskip-0.08in & \hskip-0.08in \le  & \hskip-0.08in r^{n} \|{\bf Q}{\bf x}_0-{\bf Q}{\bf u}\|_2\le r^{n} \|{\bf Q}{\bf x}_0\|_2 ,
 \vspace{-.5em}\end{eqnarray}
where $r= \max_{0\le \gamma_i<1} \gamma_i$.
This proves \eqref{maintheorem1.thm.pfeq0} and  the desired exponential convergence of the sequence ${\bf x}_n, n\ge 0$.

Taking the limit in \eqref{maintheorem1.thm.eq2} and applying the  convergence in \eqref{maintheorem1.thm.pfeq0} yields ${\bf Q}^{-2}{\bf A}^* {\bf A}  {\bf u}={\bf 0}$.
This proves that  ${\bf A}{\bf u}={\bf 0}$ and completes the proof.

\end{appendices}

\newpage


\end{document}